\documentclass[12pt,a4paper]{article}
\usepackage{amsmath,amssymb,amscd}
\usepackage{epsfig}

\begin{document}
\renewcommand{\abstractname}{Abstract}
\renewcommand{\refname}{References}

\begin{center}
{\bf STOCHASTIC JET QUENCHING IN HIGH ENERGY NUCLEAR COLLISIONS}
\end{center}

\bigskip

\begin{center}
{\bf M.R.~Kirakosyan$^{(a)}$\footnote{{\bf e-mail}: Martin.Kirakosyan@cern.ch; supported by the RFBR grant
08-02-91000}, A.V.~Leonidov$^{(a,b)\,}$\footnote{{\bf e-mail}: leonidov@lpi.ru; supported by the RFBR grants
06-02-17051 and 08-02-91000}}

\bigskip

\bigskip

(a) {\it Theoretical Physics Department, P.N.~Lebedev Physics Institute, \\ Moscow, Russia}

(b) {\it Institute of Theoretical and Experimental Physics, Moscow, Russia}

\end{center}

\bigskip

\bigskip

\begin{abstract}
Energy losses of fast color particles in random inhomogeneous color medium created in high energy nuclear
collisions are estimated.
\end{abstract}

\newpage

\section{Introduction}

Experimental \cite{RHICloss,SPSloss} and theoretical \cite{KW03,GVWZ03} studies of energy loss of fast particles provide an
extremely valuable option of a quantitative assessment of properties of dense strongly interacting medium created in
ultrarelativistic heavy ion collisions. In such high energy nuclear collisions a transformation of incident dense partonic
fluxes into a final set of inelastically produced hadrons proceeds through initial energy release via creation of physical
partonic modes followed by their transformation into a high density hadronic system that finally takes form of free-streaming
hadrons that are observed experimentally. Detailed dynamical description of high energy nuclear collisions is an extremely
difficult task, so theoretical simplifications and/or complicated Monte Carlo models are needed to describe the results of
existing experiments and predict those of future ones. The main issue addressed in such modeling is a relative role of parton
phase contribution versus the hadron phase one. One of generic features of a parton phase picture describing dynamics of
semihard partonic degrees of freedom is a pronounced turbulence-like inhomogeneity of energy release in the impact parameter
plane on the event-by-event basis \cite{GRZ97} having its origin in the powerlike nature of the cross sections of partonic
interactions. Therefore, when considering the parton phase contribution to the energy loss of fast particles, understanding
the effect of these inhomogeneities is very important.

In the present paper we shall consider this problem in the simplest setting \cite{GT91,M91} in which the energy loss is
computed, in the quasi-abelian approximation, as the work done on the incident particle by the field it induces in the medium
\cite{LL}. In this approach the origin of the losses is a nonzero imaginary part of dielectric permittivity of the medium. In
the case of random inhomogeneous medium there appears a new contribution to this imaginary part. The physical mechanism of
energy loss in the randomly inhomogeneous medium leading to an appearance of such inhomogeneity-induced imaginary part of
dielectric permittivity can be described as a scattering of the Weizsaker-Williams modes of the proper field of the incident
particle on the inhomogeneities in the medium. Depending on the stochastic properties of the medium some modes can become
resonant resulting in transition and Cherenkov radiation, see e.g. \cite{T72}. In the light of discussions on possible
observation of gluon Cherenkov radiation at RHIC, see e.g. \cite{D0708,DKLV08}, of special interest is the fact that the
inhomogeneity-induced losses are significantly amplified in the medium in which the effective permeability exceeds the
Cherenkov threshold \cite{T72}. Let us note that the problem of energy loss in the random inhomogeneous medium has deep links
\cite{K60} with that of wave propagation in random media, see e.g. the review \cite{BKRT70}.

\section{Energy loss}

The average energy losses per unit length of ultrarelativistic particle moving along the $z$ axis  are described by an
appropriate generalization \cite{GT91,M91} of the standard formula \cite{LL}:
\begin{equation}\label{genloss}
 \frac{dW}{dz} = g Q^a \hat{\bf z}\,  \langle {\bf E}^a ({\bf r},t)\rangle_{{\bf r}=\hat{\bf z}t}
\end{equation}
where $g$ is a color charge, $Q^a$ is a color spin normalized at quadratic Casimir invariants of SU(3) for
fundamental (adjoint) representations $C_{F(V)}$ , $Q^a Q^a = C_{F(V)}$, for the projectile quark or gluon
respectively and $\langle {\bf E}^a({\bf r},t)\rangle$ is a chromoelectric field created in the medium by the
propagating particle averaged over the fluctuations of random inhomogeneous medium.

In the case of inhomogeneous medium the local fluctuations of its properties are customarily parametrized by the
coordinate-dependent chromodielectric permittivity $\varepsilon^{ab}({\bf r})$. For given configuration of
$\varepsilon^{ab}({\bf r})$ the spectral component of the chromoelectric field is determined from the abelianized
equations of motion for the chromoelectric field:

\begin{equation}\label{eqE1}
 \triangle {\bf E}^a - \nabla \left( \nabla {\bf E}^a \right) + \omega^2 \, \varepsilon ({\bf r}) \, {\bf E}^a \,=
 4\pi \imath \omega j^a (\omega,{\bf r})
\end{equation}
where $j^a(\omega,{\bf r})$ is a spectral component of the external current and we have assumed a diagonal structure of
chromodielectric permittivity, $\varepsilon^{ab}({\bf r})=\varepsilon({\bf r}) \, \delta^{ab}$. In the considered case of
uniformly moving ultrarelativistic particle one has ${\bf j}^a(\omega,{\bf k})=2 \pi g Q^a \hat{\bf z} \delta (\omega - k_z)$.
Random inhomogeneities can conveniently be parametrized by explicitly identifying the non-random and random contributions to
dielectric permittivity:
\begin{equation}
 \varepsilon({\bf r}) \, = \, \varepsilon_0 \left(1+\xi ( {\bf r} )  \right)
\end{equation}
where $\xi( {\bf r} )$ is a random contribution to permittivity having zero mean $\langle {\xi(\bf r}) \rangle=0$. In what
follows we shall consider the simplest case of Gaussian ensemble so that $\xi({\bf r})$ are fully characterized by the binary
correlation function
\begin{equation}
 \langle \xi \left( {\bf r}_1 \right) \xi \left( {\bf r}_2 \right) \rangle=g\left( |{\bf r}_1-{\bf r}_2|\right)
\end{equation}
It is well known, see e.g. \cite{T72,BKRT70}, that taking into account the random inhomogeneities leads to an equation for the
average electric field containing a tensor of effective dielectric permittivity $\varepsilon_{ij}(\omega,{\bf k})$ having
nontrivial imaginary parts that depend on the statistical properties of fluctuations $\xi({\bf r})$:
\begin{equation}\label{eqE2}
\left [\varepsilon_{ij}(w, \mathbf{k})-\frac{k^2}{w^2}\left ( \delta_{i,j} - \frac{k_{i}k_{j}}{k^{2}} \right )
\right]\langle E^a_j(w, \mathbf{k}) \rangle=\frac{4 \pi}{\imath \, w}j^a_i(w, \mathbf{k})
\end{equation}
where we have introduced a notation $w=\sqrt{\varepsilon_0} \omega$. Averaging over the stochastic ensemble of
random inhomogeneities introduces an implicit dependence on its characteristics, e.g. on correlation radius $r_c$
and fluctuation magnitude $\sigma$, so that one should in fact write $\varepsilon_{ij}(w,
\mathbf{k}|\,r_c,\sigma)$. It is convenient to explicitly introduce the transverse and longitudinal components
$\varepsilon_t$ and $\varepsilon_l$:
\begin{equation}\label{epstl}
 \varepsilon_{ij} (w,{\bf k}|\,r_c,\sigma) \equiv \left(\delta_{ij}-\frac{k_ik_j}{k^2} \right)
 \varepsilon^t (w,{\bf k}|\,r_c,\sigma) + \frac{k_ik_j}{k^2} \varepsilon^l (w,{\bf k}|\,r_c,\sigma)
\end{equation}
In terms of the decomposition (\ref{epstl}) the expression for the energy losses takes, for some given
$\varepsilon_{ij}(\omega,{\bf k}|\,r_c,\sigma)$, the form
\begin{eqnarray}\label{losstl}
 \frac{dW}{dz} &=& \frac{g^2 C_{F(V)}}{2 \pi^2 } \int d^3k
 \left\{ \frac{w}{k^2} \left[ {\rm Im} \, \frac{1}{\varepsilon^l(w,{\bf k}|\,r_c,\sigma)} \right. \right. \\
 \phantom{a} &-& \left. \left. \left( w^2-k^2 \right) {\rm Im} \,
 \frac{1}{w^2 \varepsilon^t(w,{\bf k}|\,r_c,\sigma) -k^2} \right] \right\}_{w=k_z} \nonumber
\end{eqnarray}

Let us now consider the particular case of an exponential binary correlation function
\begin{equation}\label{corfun}
 g(r)=\sigma^2 e^{-r/r_c}
\end{equation}
that allows an explicit analytical computation of $\varepsilon_{ij}(\omega,{\bf k})$ to all orders in $\sigma$ in
the one-loop approximation corresponding to the regime $\sigma^2 (wr_c) \ll 1$, see Appendix. The corresponding
calculation is naturally performed in terms of the polarization tensor $\Pi_{ij}(w,{\bf k})$ related to
chromodielectric permittivity by the following relation:
\begin{equation}\label{epsPi}
 \varepsilon_{ij} (w,{\bf k}|\,r_c,\sigma) = \varepsilon_0 \left(1-\frac{1}{w^2} \Pi_{ij}(w,{\bf k}|\,r_c,\sigma) \right)
\end{equation}
The decomposition of chromodielectric permittivity (\ref{epstl}) leads to the corresponding decomposition of the
polarization tensor. The explicit expressions for the transverse and longitudinal components of the polarization
tensor $\Pi_{ij}(w,{\bf k}|\,r_c,\sigma)$ computed in the one-loop approximation read, cf. Eqs.~(\ref{APit}) and
(\ref{APil}) in the Appendix:
\begin{eqnarray}\label{Pit}
 \Pi^t (w,{\bf k}|\,r_c,\sigma)& = & \sigma^2 w^2
 \left[
    \frac{w^2}{(w+i \delta)^2-k^2}-\frac{\delta(\delta+iw)}{2k^2} \right.\\
   \phantom{a} & + & \left. \frac{\delta^2+w^2+k^2}{k^2} \frac{\delta}{k} \arctan \left( \frac{ik}{w+i\delta}\right) \nonumber
 \right]
\end{eqnarray}
and
\begin{eqnarray}\label{Pil}
 \Pi^l (w,{\bf k}|\,r_c,\sigma)& = & \sigma^2 w^2
 \left[
    1+\frac{\delta(\delta+iw)}{2k^2} \right.\\
   \phantom{a} & - & \left. \frac{\delta^2+w^2+k^2}{k^2} \frac{\delta}{k} \arctan \left( \frac{ik}{w+i\delta}\right) \nonumber
 \right]\,,
\end{eqnarray}
where $\delta=1/r_c$.
 Let us stress once again that even if the non-random chromodelectric permittivity $\varepsilon_0$ does
not have a significant imaginary part so that the corresponding energy losses determined by (\ref{losstl}) are
absent, the effective chromodielectric permittivity determined by (\ref{Pit},\ref{Pil}) has a nontrivial imaginary
part and, therefore, there arise specific energy losses directly related to the random fluctuations in the medium
in which they propagate.

Let us now estimate the magnitude of the above-described stochastic energy loss. In presenting the results it
turns out convenient to rewrite the expression for the energy loss (\ref{losstl}) in the form:
\begin{equation}\label{dimless}
 \frac{dW}{dz} =  C_{F(V)} \frac{\alpha_s }{\pi} \frac{1}{r_c^2} \left[f_t(pr_c|\,\sigma,\varepsilon_0)+f_l(pr_c|\,\sigma,\varepsilon_0) \right]
\end{equation}
where $p$ is an energy of the incident quark (gluon) and the dimensionless functions $f_t(pr_c|\,\sigma,\varepsilon_0)$ and
$f_l(pr_c|\,\sigma,\varepsilon_0)$ quantify the transverse and longitudinal energy losses correspondingly. The dependence on
$p$ appeared through cutting the momentum integration in (\ref{losstl}) at $p$ at the upper limit. The physical meaning of
this cutoff is that within the adopted scheme of calculating the energy losses we should restrict our consideration to the
soft modes in the proper field of the incident particle.

In considering transverse contribution to the energy losses it is important to distinguish between the cases of
$\varepsilon_0<1$ (below the Cherenkov threshold)  and $\varepsilon_0>1$ (above the Cherenkov threshold).

In the first case of $\varepsilon_0<1$ the transverse losses are purely stochastic so that they vanish at
vanishing fluctuation strength $\sigma$,  $f_t(pr_c|\,\sigma \to 0 ,\varepsilon_0)\to 0$. An interesting feature
of the stochastic energy loss in this regime is a dependence of the relative weight of transverse and longitudinal
contributions from the value of $\varepsilon_0$. At small $\varepsilon_0$ longitudinal losses dominate over
transverse ones while for $\varepsilon_0\lesssim1$ the transverse losses are dominant. This is illustrated in
Fig.~1 in which we plot the dependence of transverse $f_t(pr_c|\,\sigma,\varepsilon_0)$ and longitudinal
$f_l(pr_c|\,\sigma,\varepsilon_0)$ stochastic losses on $pr_c$ for $\varepsilon=0.7$, where longitudinal losses
dominate, and $\varepsilon=0.9$, where transverse losses substantially exceed the longitudinal ones.

In the second case of $\varepsilon_0>1$ the transverse losses are nonzero even at vanishing fluctuation strength
$\sigma \to 0$ due to Cherenkov radiation. In the considered simplest case of constant energy-independent
imaginary part of $\varepsilon_0$ the Cherenkov contribution to the transverse losses is very big: it is in fact
quadratically divergent at the upper limit of energy integration. Therefore to consider the stochastic
contribution to energy losses in the Cherenkov regime one has to subtract the pure Cherenkov losses from the full
transverse ones. The stochastic contribution $f^{\rm stoch}_t (pr_c|\,\sigma ,\varepsilon_0)$ to transverse energy
loss in this case is thus just $f^{\rm stoch}_t (pr_c|\,\sigma ,\varepsilon_0)=f_t(pr_c|\,\sigma ,\varepsilon_0) -
f_t(pr_c|\,\sigma=0,\varepsilon_0)$. The full transverse loss $f_t(pr_c|\,\sigma ,\varepsilon_0)$, its stochastic
component $f^{\rm stoch}_t(pr_c|\,\sigma ,\varepsilon_0)$ and the longitudinal loss $f_l(pr_c|\,\sigma
,\varepsilon_0)$ are plotted in Fig.~2 for $\varepsilon=1.1$. We see that the pure Cherenkov contribution
$f_t(pr_c|\,\sigma=0,\varepsilon_0)$ is clearly dominant while the stochastic losses $f^{\rm
stoch}_t(pr_c|\,\sigma ,\varepsilon_0)$ and $f_l(pr_c|\,\sigma ,\varepsilon_0)$ are similar to those for
$\varepsilon_0=0.9$ plotted in Fig.~1(b).

The dependence of transverse losses $f_t (pr_c|\,\sigma ,\varepsilon_0)$ and stochastic transverse  losses $f^{\rm
stoch}_t (pr_c|\,\sigma ,\varepsilon_0)$ on $\varepsilon_0$ is plotted, for $\sigma=0.2$ and $pr_c=10$, in Fig.~3.
We see that a relative weight of stochastic losses above the Cherenkov threshold is diminishing with growing
$\varepsilon_0$.

The dependence of total losses  $f_{\rm tot}(pr_c|\,\sigma,\varepsilon_0)$ on the fluctuation magnitude $\sigma$
is illustrated by plotting the dependence on $pr_c$ for several values of $\sigma$ in Fig. 4 for
$\varepsilon_0=0.7$. As expected, the losses grow with growing $\sigma$.

To get a feeling on the magnitude of the stochastic energy loss in Gev/fm we plot it, for a quark with energy up
to 20 GeV and parameter values $\alpha_s=0.2$, $r_c=0.2\,\,{\rm fm}$, $\sigma=0.2$ and $\varepsilon_0=0.9$, in
Fig. 5. We see that at high energies the transverse stochastic loss can become quite substantial and exceed the
values discussed in \cite{GT91,B82}.

Let us stress that the considered mechanism of stochastic quenching presents and {\it additional} source of energy
loss with respect to mechanisms considered earlier in the literature. It is also important to have in mind that
the above values constitute an order-of-magnitude estimate of stochastic quenching only since the above-described
calculation is not reliable at $pr_c \gg 1$, also because of the strong dependence of the calculated energy loss
on the cutoff $p$. Working out a consistent picture matching the above-described soft energy loss to the energy
loss of hard modes requires developing a microscopic approach\footnote{For an interesting attempt of a combined
treatment of ionization and collisional losses see \cite{M91}.}. This issue is a very important subject for future
analysis. Another important task is to perform a quantitative analysis of the properties of inhomogeneous gluon
medium created in ultrarelativistic nuclear collisions at LHC energies and use it as a basis for computing the
stochastic energy losses of fast quarks and gluons.  .

\section{Conclusions}

Stochastic energy loss of fast color particles in the random inhomogeneous medium was calculated to all orders in the
fluctuation magnitude in the one-loop approximation. The dependence of stochastic losses on the bare chromodielectric
permittivity $\varepsilon_0$, particle energy $p$ and fluctuation strength $\sigma$ was analyzed. The analysis has shown that
stochastic energy loss can be of phenomenological importance. The strong dependence of the results on the value of ultraviolet
cutoff indicates a necessity of developing a consistent microscopic approach allowing a unified treatment of energy losses due
to soft and hard modes enabling a reliable quantitative calculation of the corresponding stochastic contribution. It is also
of interest to consider in more details an interrelation between the above-considered mechanism of energy loss and effects of
generic stochastic acceleration mechanism in random color fields considered in \cite{L95}.

\begin{center}
{\bf Acknowledgements}
\end{center}

The authors are indebted to I.M. Dremin, S.M. Apenko, V.V. Losyakov and A.A. Rukhadze for useful discussions. We are also
grateful to I.M. Dremin for reading the manuscript and suggesting improvements.

\section*{Appendix: Averaged vector field in statistically homogeneous medium.}

To compute an expression for the disorder-averaged chromoelectric field determining the mean energy loss we have
to write down a solution of the equation (\ref{eqE1}) for some given random profile $\xi({\bf r})$ and then
average it over the Gaussian ensemble of random inhomogeneities with the two-point correlator (\ref{corfun}). The
first step is writing down an expression for the Green function ${\bf G}(w,{\bf r} |\,{\bf n},{\bf r}_0)$ obeying
the following equation\footnote{In this appendix we omit the color indices. In the considered abelian
approximation the answer corresponds to a given mode of the chromoelectric field.}:
\begin{equation}\label{egf1}
 \triangle{\bf G} -\nabla (\nabla{\bf G})+w^{2}{\bf G}=
 -w^2 \xi ({\bf r}){\bf G}+{\bf n}\delta ({\bf r}-{\bf r_{0}})\, ,
\end{equation}
where ${\bf n}$ is a fixed unit three-vector and ${\bf r}_0$ is a fixed spatial point. In momentum space the
equation (\ref{egf1}) reads
\begin{equation}\label{egf2}
 \left[ (w^2-{\bf k}^2) \delta_{i j}+k_{i} k_{j} \right] G_j({\bf k}) \, = \,
 -w^2 ( \hat{ \xi } G_{i} ) ({\bf k})+ \frac{1}{(2\pi)^3} n_{i} {\rm e}^{ -\imath {\bf k} {\bf r_0} } \, ,
\end{equation}
where
\begin{equation}
 (\hat{\xi}{\mathbf G})({\mathbf k})=
 \frac{1}{(2\pi)^3} \int d^3 {\bf r} d^3 {\bf k'} \xi({\mathbf r}){\mathbf G}({\bf k'})
 {\rm e}^{\imath({\bf k'}-{\mathbf k}){\bf r}}
\end{equation}

Let us consider an iterative solution of (\ref{egf1}):
\begin{eqnarray}\label{itsol}
 G_{i}({w,\bf k})&=&\frac{1}{(2 \pi)^3} G^0_{i j} n_j {\rm e}^{ -i{\bf k}{\bf r_0}}-
 \frac{1}{(2\pi)^6}w^2G^0_{i j_{1}}(\hat{\xi}A_{j_1 j}n_{j} {\rm e}^{-\imath {\bf kr_0}}) \nonumber \\
 \,&+& \frac{1}{(2\pi)^9}w^4G^0_{i j_{1}}(\hat{\xi}G^0_{j_1 j_2}(\hat{\xi}G^0_{j_2 j}n_{j}
 {\rm e}^{-\imath{\bf kr_0}})) +\ldots \, ,
\end{eqnarray}
where $G^0_{i j}(w,{\bf k})$ is a free propagator
\begin{equation*}
 G^0_{ij}=\frac{\delta_{ij}}{w^2-k^2}+\frac{k_{i}k_{j}}{w^2 (k^2-w^2)}
\end{equation*}

After averaging equation (\ref{itsol}) over the gaussian ensemble of fluctuations $\{ \xi({\bf r}) \}$ with the
two-point correlator (\ref{corfun}) we arrive at the algebraic Dyson equation for the averaged Green function
$\langle G_i (w,{\bf k}|\,r_c,\sigma,{\bf n}) \rangle$:
\begin{equation}\label{dyseq}
 \langle G_i \rangle = \frac{1}{ (2\pi)^3} G^0_{i j} n_j {\rm e}^{-\imath{\bf k}{\bf r_0}}+G^0_{i k}\Pi_{k j}
 \langle G \rangle_j
\end{equation}
so that
\begin{equation}\label{dyseqsol}
 \langle {\bf G}(w,{\bf k}|\,r_c,\sigma,{\bf n}) \rangle =
 \frac{G^0(w,{\bf k})}{I-G^0(w,{\bf k}) \Pi(w,{\bf k}|\,r_c,\sigma)} \frac{{\bf n}}{(2 \pi)^3} {\rm e}^{-\imath{\bf k}{\bf r_0}}
\end{equation}
The averaged Green function (\ref{dyseqsol}) allows to compute the averaged chromoelectric field for an arbitrary
external current $j^a({\bf r},t)$.

 In the limit $\sigma^2 (kr_c) \ll 1$ one can keep the one-loop approximation
for the polarization operator T$\Pi(w,{\bf k}|\,r_c,\sigma)$:
\begin{equation}
 \Pi_{i j}(w,{\bf k}|\,r_c,\sigma) \approx w^4 \int d^{3}k_1 g(|\,{\bf k_1}-{\bf k}|,r_c,\sigma)G^0_{i j}(w,{\bf k_1})
\end{equation}
which can be computed analytically. Explicit expressions for its transverse and longitudinal components
$\Pi^t_{ij} (w,{\bf k})$ and $\Pi^l_{ij} (w,{\bf k})$ take the form \cite{KL08}:
\begin{eqnarray}\label{APit}
 \Pi^t (w,{\bf k}|\,r_c,\sigma)& = & \sigma^2 w^2
 \left[
    \frac{w^2}{(w+i \delta)^2-k^2}-\frac{\delta(\delta+iw)}{2k^2} \right.\\
   \phantom{a} & + & \left. \frac{\delta^2+w^2+k^2}{k^2} \frac{\delta}{k} \arctan \left( \frac{ik}{w+i\delta}\right) \nonumber
 \right]
\end{eqnarray}
and
\begin{eqnarray}\label{APil}
 \Pi^l (w,{\bf k}|\,r_c,\sigma)& = & \sigma^2 w^2
 \left[
    1+\frac{\delta(\delta+iw)}{2k^2} \right.\\
   \phantom{a} & - & \left. \frac{\delta^2+w^2+k^2}{k^2} \frac{\delta}{k} \arctan \left( \frac{ik}{w+i\delta}\right) \nonumber
 \right]\,,
\end{eqnarray}
where $\delta=1/r_c$.

\begin{figure}[ht]
\begin{center}
 \epsfig{file=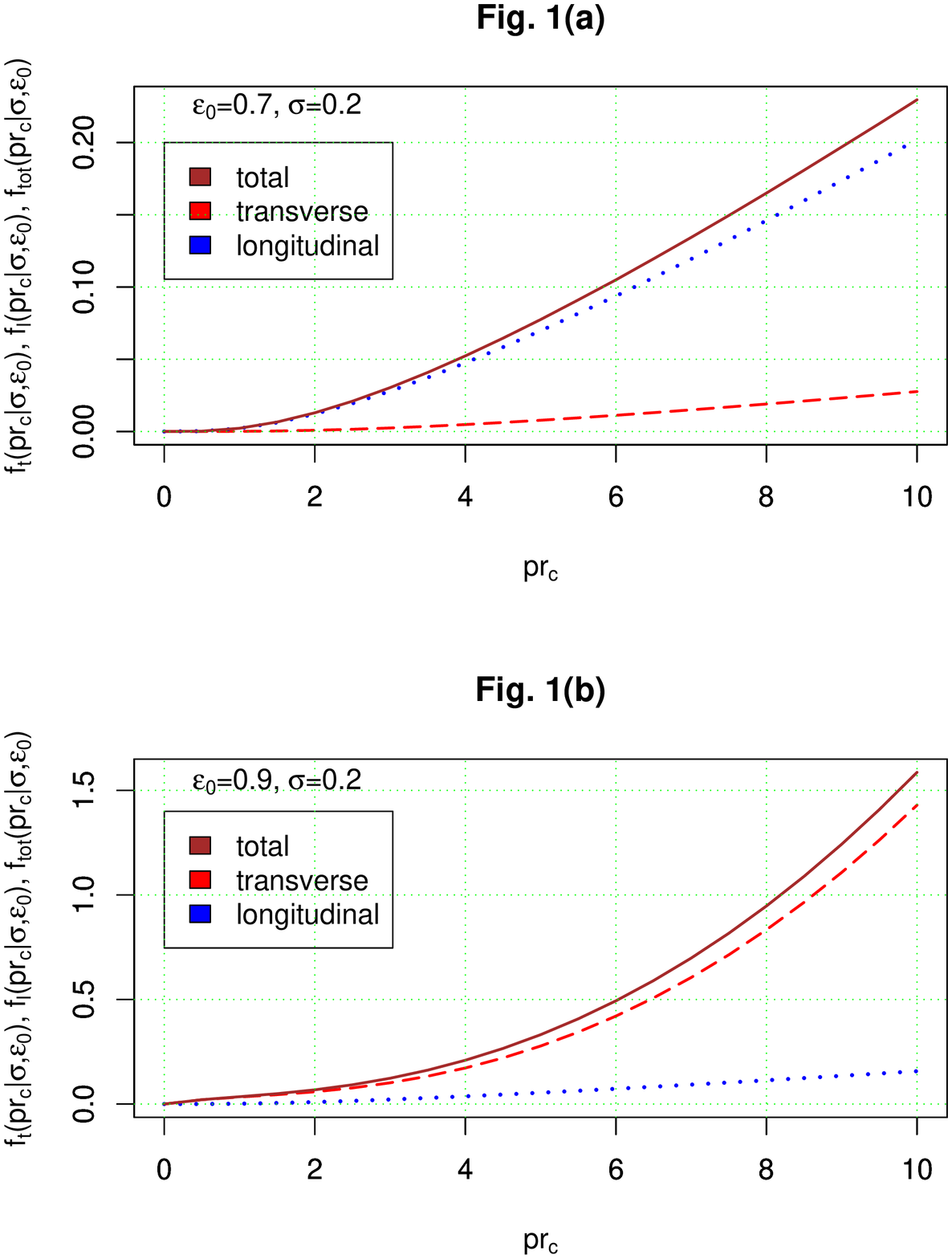,height=18cm,width=17cm}
 \caption{Transverse losses $f_t(pr_c|\,\sigma,\varepsilon_0)$ (red, dashed), longitudinal losses $f_l(pr_c|\,\sigma,\varepsilon_0)$
(blue, dotted) and total losses $f_{\rm tot}(pr_c|\,\sigma,\varepsilon_0)$ (brown, solid), $\sigma=0.2$: (a) for
$\varepsilon_0=0.7$; (b) for $\varepsilon_0=0.9$.}
\end{center}
\end{figure}

\begin{figure}[ht]
\begin{center}
 \epsfig{file=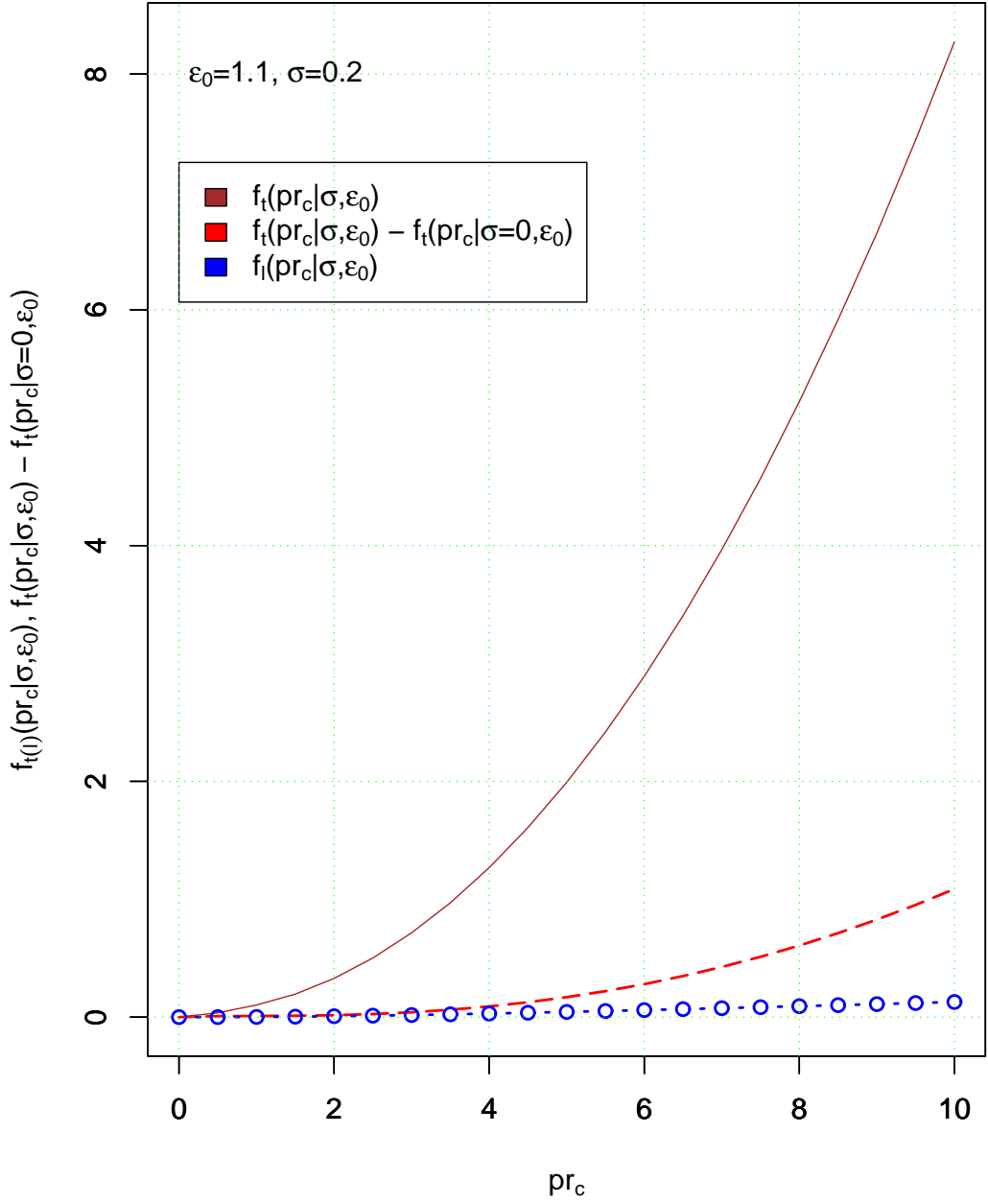,height=19cm}
 \caption{Full transverse losses $f_t(pr_c|\,\sigma,\varepsilon_0)$ (brown, solid), stochastic contribution to
 transverse losses $f^{\rm stoch}_t(pr_c|\,\sigma,\varepsilon_0)=f_t(pr_c|\,\sigma,\varepsilon_0)-f_t(pr_c|\,\sigma=0,\varepsilon_0)$ (red, dashed) and
 longitudinal losses $f_l(pr_c|\,\sigma,\varepsilon_0)$ (blue, dotted), $\sigma=0.2$, $\varepsilon_0=1.1$.}
\end{center}
\end{figure}

\begin{figure}[ht]
\begin{center}
 \epsfig{file=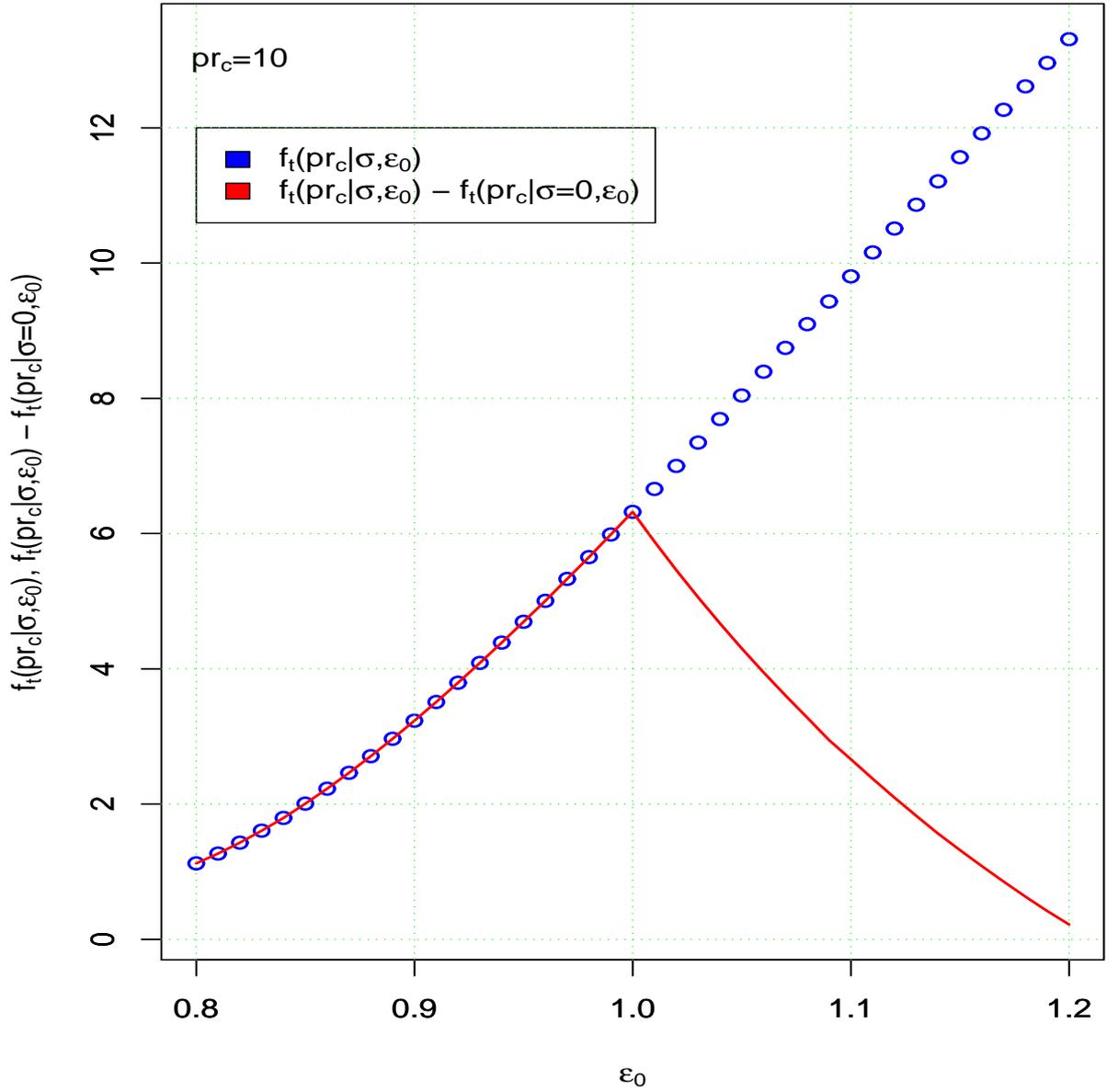,height=18cm,width=17cm}
 \caption{Dependence of full transverse losses $f_t(pr_c|\,\sigma,\varepsilon_0)$ (blue, dotted) and stochastic contribution to
 transverse losses $f^{\rm stoch}_t(pr_c|\,\sigma,\varepsilon_0)=f_t(pr_c|\,\sigma,\varepsilon_0)-f_t(pr_c|\,\sigma=0,\varepsilon_0)$ (red, solid) on the bare
 dielectric permittivity $\varepsilon_0$, $\sigma=0.2$, $pr_c=10$.}
\end{center}
\end{figure}

\begin{figure}[ht]
\begin{center}
 \epsfig{file=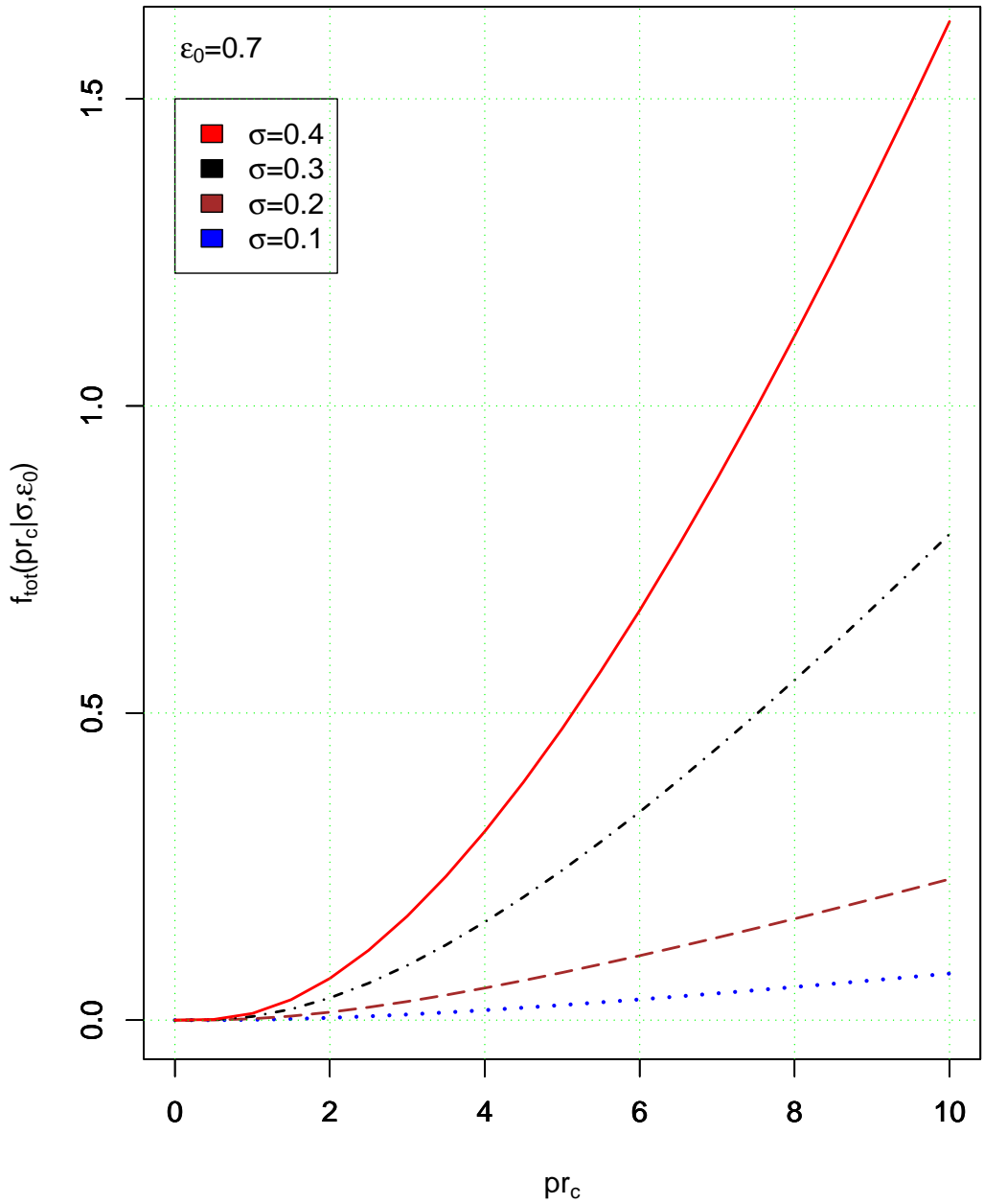,height=19cm}
 \caption{Dependence of total losses $f_{\rm tot}(pr_c|\,\sigma,\varepsilon_0)$ on $pr_c$ for $\sigma=0.1$ (blue, dotted);
$\sigma=0.2$ (brown, dashed);  $\sigma=0.3$ (black, dashed-dotted); $\sigma=0.4$ (red, solid) for
$\varepsilon_0=0.7$. }
\end{center}
\end{figure}

\begin{figure}[ht]
\begin{center}
 \epsfig{file=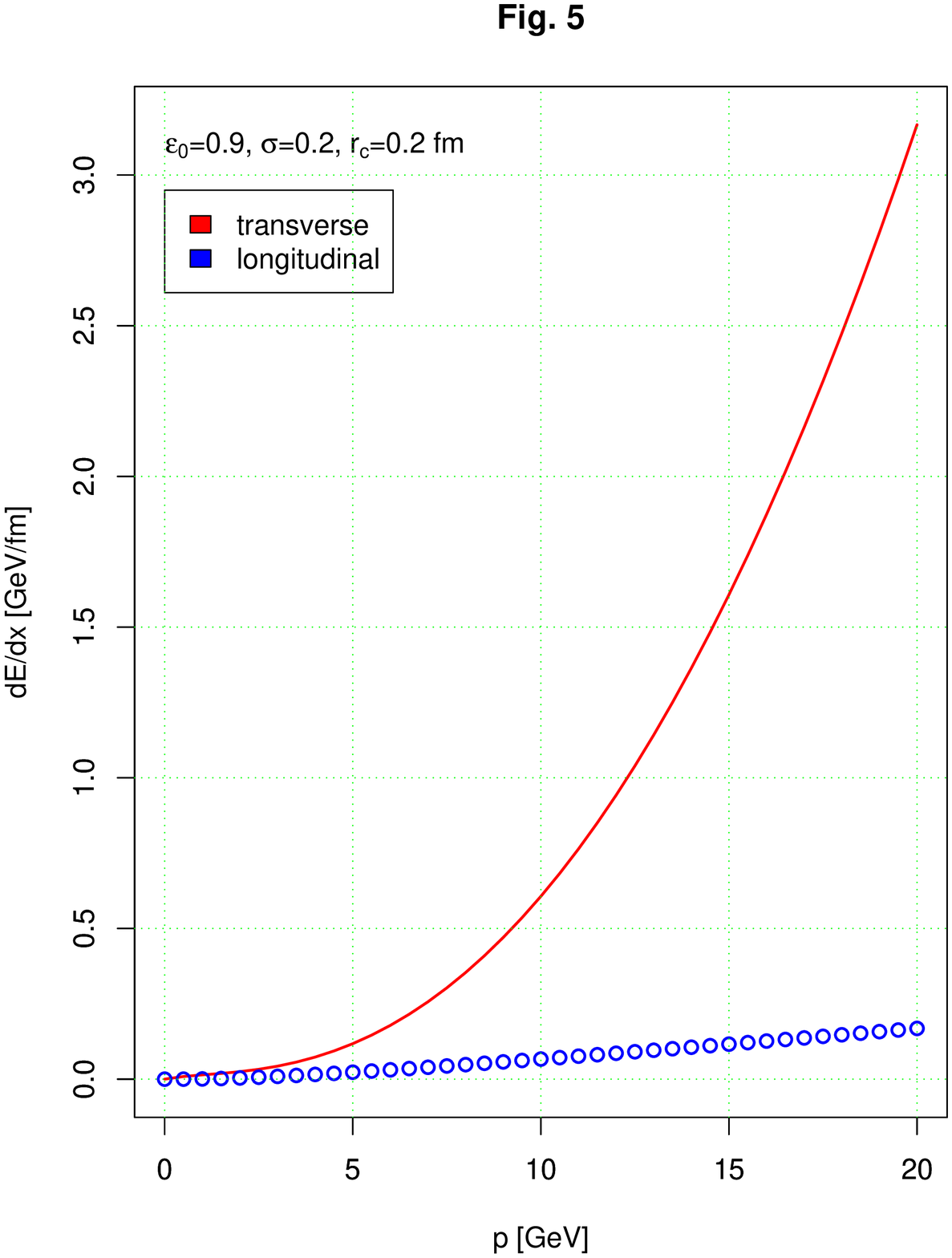,height=19cm}
 \caption{Energy dependence of quark transverse (red, solid) and longitudinal (blue, dotted) stochastic energy losses,
 $r_c=0.2\,{\rm fm}$, $\sigma=0.2$, $\varepsilon_0=0.9$. }
\end{center}
\end{figure}


\newpage
\begin{thebibliography}{99}


\bibitem{RHICloss}
K. Adcox et al. [PHENIX Collaboration], {\it Phys. Rev. Lett.}\ {\bf 88} (2002), 022301\\
C. Adler et al. [STAR Collaboration], {\it Phys. Rev. Lett.}\ {\bf 89} (2002), 202301\\
C. Adler et al. [STAR Collaboration], {\it Phys. Rev. Lett.}\ {\bf 90} (2003), 082302\\
S.S. Adler et al. [PHENIX Collaboration], {\it Phys. Rev. Lett.}\ {\bf 91} (2003), 072301\\
J. Adams et al. [STAR Collaboration], {\it Phys. Rev. Lett.}\ {\bf 91} (2003), 172302

\bibitem{SPSloss}
C. Alt et al. [NA49 Collaboration], {\it Nucl. Phys.}\ {\bf A774} (2006), 473

\bibitem{KW03}
A. Kovner, U. Wiedemann, "Gluon radiation and parton energy loss", in Quark Gluon Plasma 3, Editors: R.C. Hwa and
X.N. Wang, World Scientific, Singapore, 2003, 192; ArXiv: hep-ph/0304151.

\bibitem{GVWZ03}
M. Gyulassy, I. Vitev, X.-N. Wang, B.-W. Zhang, "Jet quenching and radiative energy loss in dense nuclear matter",
in Quark Gluon Plasma 3, Editors: R.C. Hwa and X.N. Wang, World Scientific, Singapore, 2003, 123; ArXiv:
nucl-th/0302077.

\bibitem{GRZ97}
M. Gyulassy, D.H. Rischke, B. Zhang, {\it Nucl. Phys.}\ {\bf A613} (1997), 397

\bibitem{GT91}
M. Gyulassy, M. Thoma, {\it Nucl.Phys.}\, {\bf B351} (1991), 491

\bibitem{M91}
S. Mrowczynski, {\it Phys. Lett.}\, {\bf B269} (1991), 383

\bibitem{LL}
L.D. Landau, E.M. Lifshitz, L.P. Pitaevski, "Electrodynamics of continuous media", Pergamon Press, Oxford, 1984

\bibitem{T72}
V.V. Tamoykin, {\it Astrophysics and Space Science}, {\bf 16} (1972), 120

\bibitem{D0708}
I.M. Dremin, {\it Int. J. Mod. Phys.}\, {\bf A22} (2007) 3087; {\it J. Phys.} {\bf G35} (2008) 054001

\bibitem{DKLV08}
I.M. Dremin, M.R. Kirakosyan, A.V. Leonidov, A.V. Vinogradov, "Cherenkov Glue in Opaque Nuclear Medium",
arXiv:0809.2472 [hep-ph]

\bibitem{K60}
S.P. Kapitsa, {\it ZHETP} {\bf 39} (1960), 1367

\bibitem{BKRT70}
Yu.N. Barabanenkov, Yu.A. Kravtsov, S.M. Rytov, V.I. Tatarsky, {\it UFN}\ {\bf 102} (1970), 3 (in Russian)

\bibitem{B82}
J.D. Bjorken, {\it Energy Loss of Energetic Partons in Quark-Gluon Plasma: Possible Extinction of High $p_T$ Jets
in Hadron-Hadron Collisions}, preprint FERMILAB-Pub-82/59-THY

\bibitem{KL08}
M.R. Kirakosyan, A.V. Leonidov, "Energy Loss in Stochastic Abelian Medium", arXiv:0809.2179 [hep-ph]

\bibitem{L95}
A. Leonidov, {\it Zeit. Phys.}\ {\bf C66} (1995), 263

\end{thebibliography}
\end{document}